\documentclass{nature}
\usepackage{amsmath}
\usepackage{graphicx}
\usepackage{txfonts}
\usepackage[nointegrals]{wasysym}
\usepackage{xcolor}
\usepackage{ccaption}
\usepackage{hyperref}

\title{Potential long-term habitable conditions on planets with primordial H-He atmospheres}

\author{Marit Mol Lous $^{1,2}$, Ravit Helled$^2$ \& Christoph Mordasini$^1$}

\begin{document}
\maketitle

\begin{affiliations}
 \item Physikalisches Institut, Universit\"at Bern, Gesellschaftstrasse 6, 3012 Bern, Switzerland
 \item Institute for Computational Science, University of Z\"urich, Winterthurerstrasse 190, CH-8057 Z\"urich, Switzerland
\end{affiliations}

\begin{abstract}
Cold super-Earths which retain their primordial, H-He dominated atmosphere could have surfaces that are warm enough to host liquid water. This would be due to the collision induced absorption (CIA) of infra-red light by hydrogen, which increases with pressure. However, the long-term potential for habitability of such planet has not been explored yet. Here we investigate the duration of this potential exotic habitability by simulating planets of different core masses, envelope masses and semi-major axes. We find that terrestrial and super-Earth planets with masses of $\sim$1 - 10$M_{\oplus}$ can maintain temperate surface conditions up to 5 - 8 Gyr at radial distances larger than $\sim $2 AU. The required envelope masses are $\sim 10^{-4} \, M_{\oplus}$ (which is 2 orders of magnitude more massive than Earth's), but can be an order of magnitude smaller (when close-in) or larger (when far out). This result suggests that the concept of planetary habitability should be revisited and made more inclusive with respect to the classical definition.
\end{abstract}

The emergence of life as we know it is generally understood to require three building blocks: an energy source, access to nutrition and the presence of liquid water \cite{Lammer_2009, Irwin_2020}. Apart from these requirements, it remains unknown to what extent conditions on other planets need to resemble Earth in order to be habitable. Since the possibility of in-situ solar system exploration, habitable conditions are considered on planets and moons that have liquid water oceans underneath a layer of ice \cite{Hussmann_2006}. These habitats would have to be inhabited by organisms that are adapted to high pressures and use a metabolism independent of photosynthesis. Such organisms have already been found on Earth: while today the majority of Earth’s biomass is concentrated on its surface (due to complex photosynthetic organisms such as land plants\cite{Bar-On_2018}), the subsurface biomass likely outweighed the surface biosphere for most of geologic history \cite{McMahon_2018}. Life has adapted to many other, relatively extreme, environments such as the depths of the ocean at kbar pressures \cite{Edwards_2012}, although it is unknown how many of these organisms live independently from life on the surface. In the search for life on extraterrestrial planets it needs to be considered that life might manifest and thrive under conditions that would be considered extreme on Earth.\\ 
Observations of exoplanets have shown that the solar system might not be a standard planetary system. This was initially implied after the first detections of Hot Jupiters \cite{Mayor_1995} (although these turned out later to be rather rare but seemingly prevalent due to a detection bias) and further confirmed by the fact that super-Earths, absent in the solar system, are very common \cite{Fulton_2017, Wu_2019, Otegi_2020}.
To date, super-Earths have only been observed at short orbits, but planet formation models indicate they are also prevalent at large radial distances \cite{Emsenhuber_2021b} where they could retain a non-negligible H-He envelope, which is the original atmosphere accreted from the protoplanetary disk \cite{Owen_2020, Rogers_2021}. At several AU away from the host star, stellar radiation is insufficient to thermally evaporate a significant amount of the atmospheric H-He \cite{Lammer_2014}.\\
The primordial atmosphere, dominated by hydrogen and helium, would have insufficient greenhouse gases that are significant on Earth, such as $\text{CO}_{2}$ or methane. However, if the atmosphere is massive enough, $\text{H}_{2}$ will act as a greenhouse gas. At sufficient pressures the $\text{H}_{2}$ molecules undergo enough collisions to create a dipole moment, causing them to absorb the infrared radiation coming from the planet; this is known as \textit{collision induced absorption} (CIA) \cite{Borysow_2002, Frommhold_2006}. It could raise the surface temperature enough to allow a liquid water ocean \cite{Stevenson_1999, Pierrehumbert_2011, Wordsworth_2012}.\\
Past analytical work has shown that in the case of a massive atmosphere an intrinsic heat source is sufficient to warm the planetary surface, possibly even enabling unbound planets to be temperate \cite{Stevenson_1999}. Alternatively, planets with a smaller envelope can have stellar irradiation as the sole energy source, assuming that the accreted envelope is neither too massive/small for the given equilibrium temperature \cite{Pierrehumbert_2011}. Recently Madhusudhan et al.\cite{Madhusudhan_2021} investigated the mass-radius relationship of 'Hycean worlds', planets with a liquid water layer above a rocky core and below H-He primordial atmosphere. Similar to previous studies \cite{Pierrehumbert_2011, Seager_2013}, they found that liquid water can persist at a wide range of equilibrium temperatures. Therefore, the range of orbital distances where liquid surface water can exist (the so-called habitable zone \cite{Kasting_1993, Kopparapu_2013, Kopparapu_2014}) is wider than in the case of Earth-like planets.\\
These past studies concentrated on the static properties necessary for liquid water. However, liquid water conditions on these cold planets could be a transient phase \cite{Wordsworth_2012}. It is important to consider atmospheric escape since this can give limits to the separations at which H-He envelopes allow  the existence and stability of liquid water. Here, we study the temporal evolution in particular to address the duration of a primordial H-He habitat.\\
\\
We investigate how long surface liquid water could be permitted using an evolution model and define the duration of adequate surface pressure-temperature conditions as $\tau_{\text{lqw}}$. First  we assume that the core mass, envelope mass and semi-major axis remain constant in time. We run a large grid of evolution models and vary these three parameters. During the planetary evolution the equilibrium temperature changes due to the evolution of the host-star, which is modelled as sun-like \cite{Baraffe_2015}. The planet's intrinsic luminosity declines as the  envelope and interior cool down and radiogenic components decay. For a given core mass we run 1144 models, with the semi-major axis varied from 1 AU to 100 AU and the envelope mass from $10^{-6}$ to $10^{-1.8}$ $M_{\oplus}$.\\
Figure \ref{fig:duration_noloss} shows these grids for models with core masses of 1.5, 3 and 8 $M_{\oplus}$. Every grid-point represents a planet with a given core mass, envelope mass, and semi-major axis. The evolution models are run for 8 Gyr (see Method for discussion). Colored, circular grid-points represent $\tau_{\text{lqw}}$ of at least 10 Myr. Planets with the longest duration, $\tau_{\text{lqw}} \, >$  5 Gyr, are indicated in yellow. Within the range of considered core masses, the values of $\tau_{\text{lqw}}$ have the same trends with respect to envelope masses and semi-major axes. At orbits within $\sim$10 AU long-term $\tau_{\text{lqw}}$, indicated by the yellow data-points, are distributed over envelope masses of $10^{-4}$ to $10^{-6}$. For envelopes smaller than $\sim 10^{-5}$ the surface temperatures are temperate, similar to Earth's. The stellar radiation is expected to be the dominant source of energy and there is a scaling relation between the received flux (through the semi-major axis) and the envelope mass for long-term $\tau_{\text{lqw}}$. Beyond $\sim$10 AU liquid water conditions follow the cooling of the interior. Planets with small envelope masses have liquid water conditions relatively early while planets with more massive envelope reach liquid water conditions later in the evolution.\\
In addition to $\tau_{\text{lqw}}$ we show a case where an additional constraint is imposed: the surface temperature must remain below 400 K. This is based on the temperature limit for life on Earth \cite{Takai_2008}. The bottom right panel of Figure \ref{fig:duration_noloss} shows the results for this case for a planet with a 3 $M_{\oplus}$ core. \\

\subsection{Including atmospheric loss}
\label{subsec:atms_loss_results}
The results presented so far do not include the effect of atmospheric loss. This is important for small planets close to their host stars since they are much more vulnerable to stellar radiation \cite{Pierrehumbert_2011, Wordsworth_2012, Lammer_2014}. We apply the evaporation model discussed in the Methods to study which planets in the previous result would be stable against thermal atmosphere loss.\\
Figure \ref{fig:loss_cases} in the methods shows how the envelope mass changes with time with hydrodynamic escape (top) and Jeans loss (bottom). The Jeans escape model only has a significant effect on the mass of envelopes around small core masses of 1.5 $M_{\oplus}$. For the Jeans model we show the highest considered exosphere temperature of 2,000 K . This is relatively high as exosphere temperatures for H-He compositions are estimated at $\sim$1000 K \cite{Tian_2015}. Since a higher exosphere temperature leads to a higher mass loss rate, we consider this as an upper limit to Jeans escape. Lower values of 300 K  and 1000 K  were also considered, but we find almost no effect on the envelope's mass evolution. Even for the cases in which Jeans escape removes some of the planetary envelope, we still find a similar $\tau_{\text{lqw}}$, suggesting that the results presented in Figure ~\ref{fig:duration_noloss} are robust if escape occurs in the Jeans regime. 
The XUV-driven hydrodynamic model, however, can significantly affect the evolution of small close-in planets over a wide range of parameters, as shown in Figure \ref{fig:loss_cases}. \\
The results when the hydrodynamic escape model is included are shown in Figure \ref{fig:evol_4cores}. In this case we find that there are no long-term liquid water conditions possible on planets with a primordial atmosphere within 2 AU. Madhusudhan et al. 2021\cite{Madhusudhan_2021} found that for planets around sun-like stars, liquid water conditions are allowed at a distance of $\sim$1 AU. We find that the pressures required for liquid water conditions between 1 and 2 AU are too low to be resistant against atmospheric escape, assuming that the planet does not migrate at a late evolutionary stage.\\

\subsection{Unbound Planetary Mass Objects}
\label{subsec:rogue}
\begin{figure}
    \centering
    \includegraphics[width=1.0\textwidth]{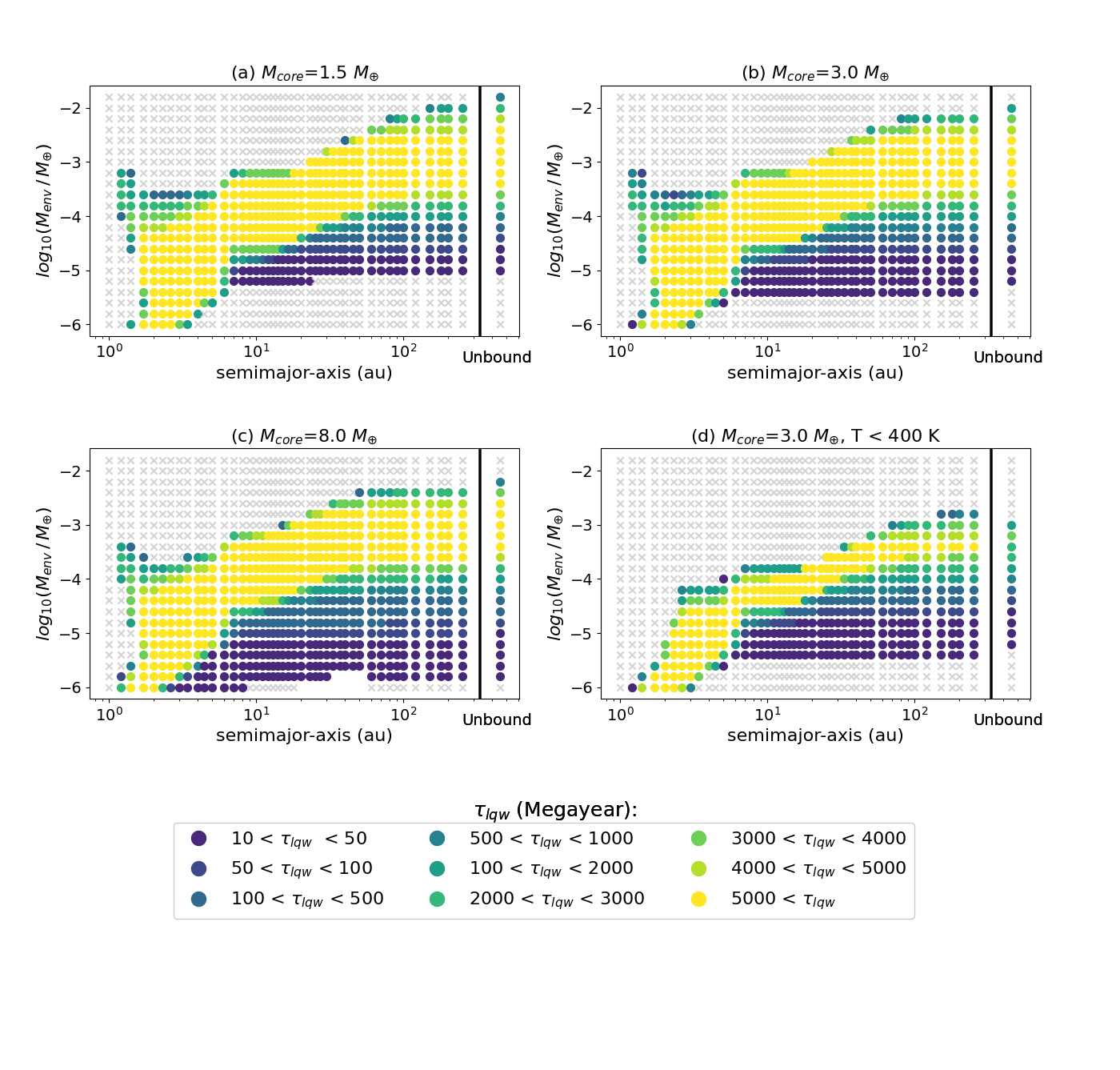}
    \vspace*{-30mm}
    \caption{Duration of liquid water conditions for planets at a wide range of semi-major axis (1 AU to 100 AU) and envelope masses ($10^{-1.8}$ to $10^{-6}$ $M_{\oplus}$). Planets receive insolation based on the luminosity evolution of a sun-like star. Panels a-c correspond to core masses of 1.5, 3 and 8 $M_{\oplus}$ respectively. The duration of the total evolution is 8 Gyr. The color of a grid-point indicates how long there were continuous surface pressures and temperatures allowing liquid water, $\tau_{\text{lqw}}$. These range from 10 Myr (purple) to over 5 Gyr (yellow). Grey crosses correspond to cases with no liquid water conditions lasting longer than 10 Myr. Atmospheric loss is not considered in these simulations. Panel d shows results for planets with a core mass of 3 $M_{\oplus}$, but including the constraint that the surface temperature must remain between 270 and 400 K. Every panel contains an 'unbound' case where the distance is set to 10$^{6}$ AU and solar insolation has become negligible.}
    \label{fig:duration_noloss}
\end{figure}

\begin{figure*}
    \centering
    \includegraphics[width=1.\textwidth]{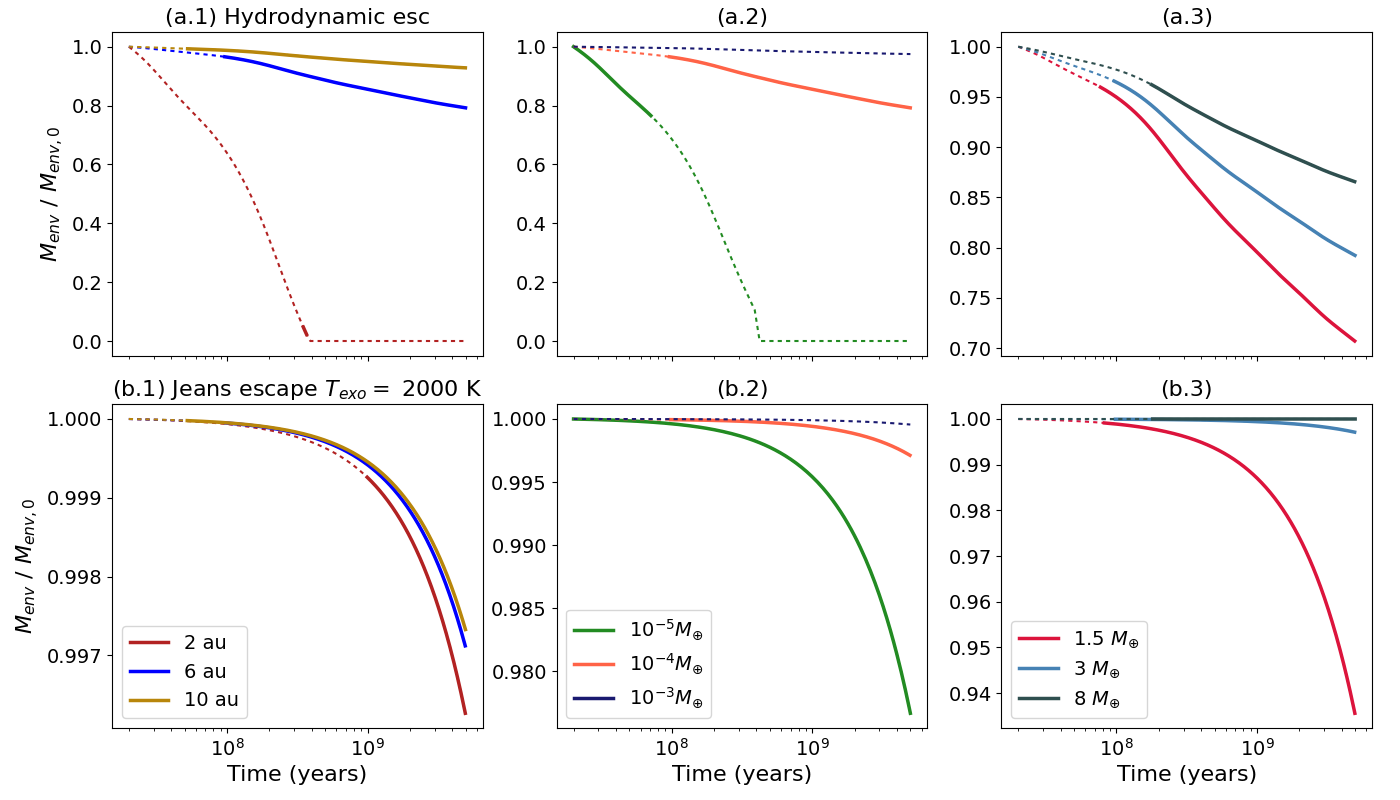}
    \caption{The evolution of the mass of the simulated envelope for two different atmosphere loss models: Hydrodynamic escape (top) and Jeans escape (bottom). The default values of the model are a core mass of 3 $M_{\oplus}$ and envelope mass of $10^{-4} \, M_{\oplus}$ at a distance of 6 AU. The top panel shows how the envelope loss is affected by different semi-major axis, the middle panel shows the effect of different envelope masses and the bottom panel shows the effect of different core masses. Dashed lines show when a layer of water between the envelope and core would not be in the liquid phase, solid lines show when it would. Hydrodynamical escape can significantly reduce or completely evaporate the envelope of some planets, which influences the duration of liquid water conditions. The Jeans escape model, even when a high exosphere temperature is chosen of 2000 K , shows only a somewhat significant loss for a small core mass of 1.5 $M_{\oplus}$.}
    
    \label{fig:loss_cases}
\end{figure*}

\begin{figure*}
    \centering
    \includegraphics[width=1.\textwidth]{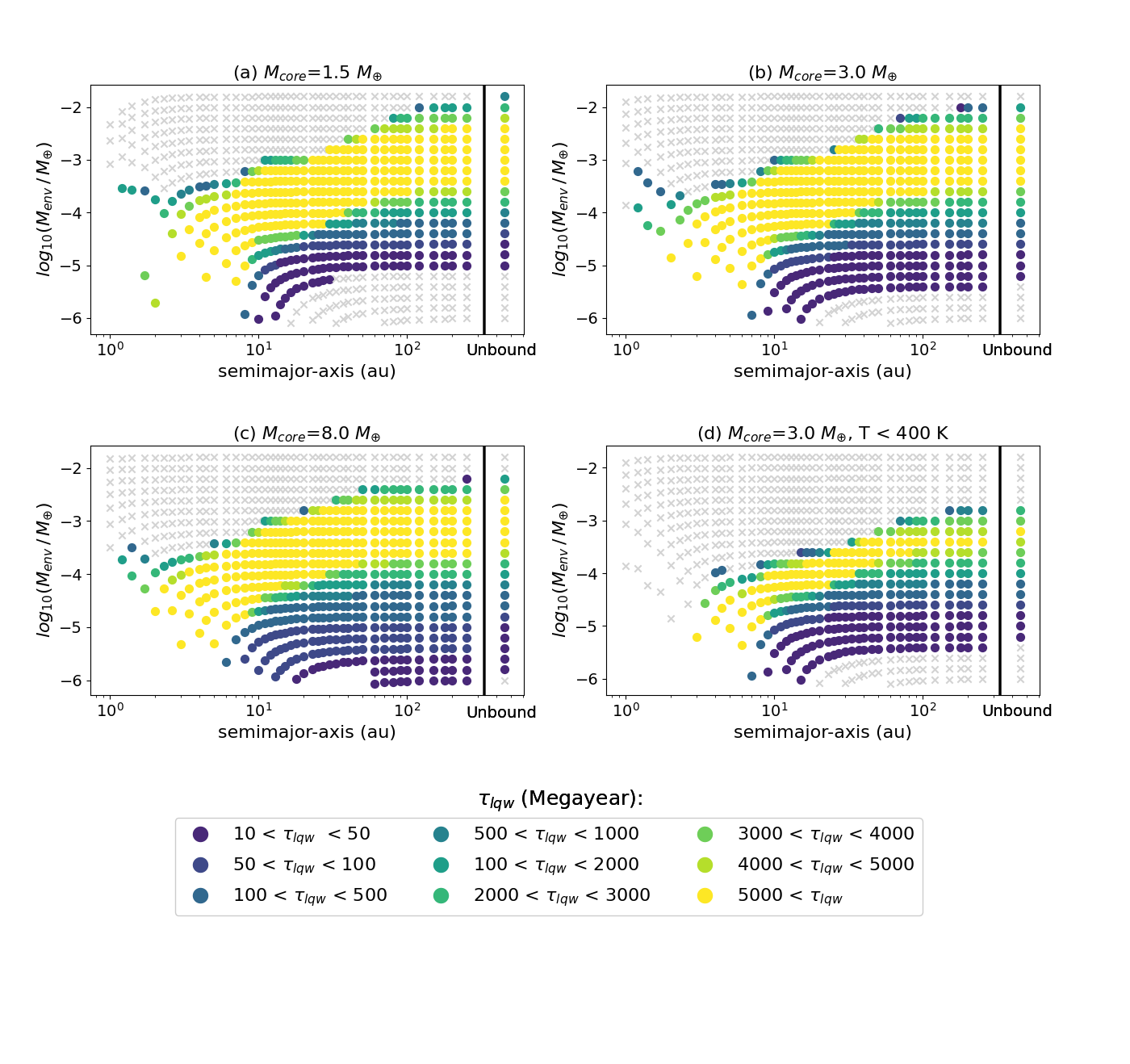}
    \caption{Same as Figure \ref{fig:duration_noloss}, but when including the effect of hydrodynamic atmospheric photo-evaporation. The y-axis shows the envelope mass after 8 Gyr. Escape inhibits liquid water conditions by removing the atmosphere for close-in planets with low initial envelope masses. Lower core masses are more affected.}
    \label{fig:evol_4cores}
\end{figure*}

\begin{figure}
    \centering
    \includegraphics[width=1.\textwidth]{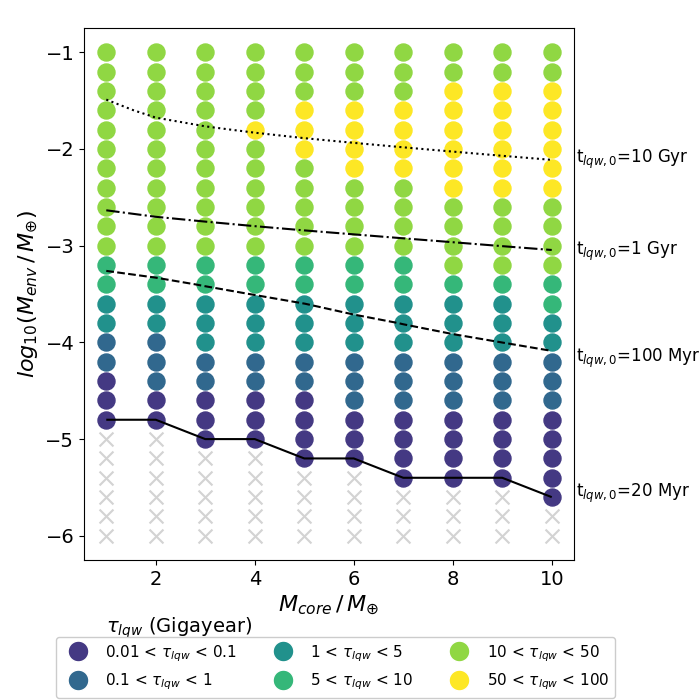}
    \caption{Duration of liquid water conditions ($\tau_{\text{lqw}}$) on unbound  planets. The planets are simulated with different core masses and envelope masses. The longest duration simulated was 84 billion years for a 10 $M_{\oplus}$ core and a 0.01 $M_{\oplus}$ envelope. If planets with a primordial atmosphere can host liquid water the duration can be much longer on unbound planets, since the internal heat source can evolve slower than the host-star. Contour lines indicate the start of liquid water conditions for planets with $\tau_{\text{lqw}} > $ 100 Myr.}
    \label{fig:cold_duration}
\end{figure}

The cases labelled as 'Unbound' shown in Figure \ref{fig:duration_noloss} and \ref{fig:evol_4cores} correspond to planets that are detached from their host star. Planets can be ejected after formation by gravitational interaction  \cite{Lissauer_1987}. Such unbound planets might be very common \cite{Strigari_2012}. Without the influence of a host-star, these planets are not affected by post main-sequence evolution, which might terminate the duration of habitable conditions for bound planets. Instead, the habitable conditions would end when an internal heat source can no longer provide enough energy. Note that unbound planets are no longer protected by the stellar heliosphere and therefore galactic cosmic-rays and gamma ray could affect the planet's structure and its potential habitability \cite{Avila_2021}.
\\
We simulate such cold planets on a longer timescale and investigate whether $\tau_{\text{lqw}}$ persists beyond the life-time of a sun-like star. The results  are shown in Figure \ref{fig:cold_duration}. The simulations are performed up to 100 Gyr, when most of the radioactive nuclei have decayed and the intrinsic luminosity is nearly zero. The different results are caused by the longer integration time than for the unbound cases in Figures \ref{fig:duration_noloss} and \ref{fig:evol_4cores}.\\
We find that many unbound planets are too hot shortly after their formation to host liquid water but do have the right conditions at later times and for long $\tau_{\text{lqw}}$. In this case the core mass is an important factor for $\tau_{\text{lqw}}$, since we assume that the radiogenic heat source scales with core mass. The envelope's mass influences the planetary cooling efficiency and therefore has a smaller, but still significant, effect on $\tau_{\text{lqw}}$. Planets with cores more massive than 5 $M_{\oplus}$ can have liquid water conditions lasting over 50 Gyr for envelopes with masses of $\sim$0.01 $M_{\oplus}$. The longest duration for liquid-water is found to be 84 Gyr for a planet with 10 $M_{\oplus}$ core and 0.01 $M_{\oplus}$ envelope. The current set-up of our intrinsic luminosity model limits $\tau_{\text{lqw}}$ of unbound planets by the half-life of the radio-active components in the core, assuming a chondritic abundance. A potential planet with radioactive components with a longer half-life could therefore have an even longer $\tau_{\text{lqw}}$.
Many of the inferred $\tau_{\text{lqw}}$ values for unbound planets are much longer than the age of the universe. It should be noted, however, that it takes $\sim$10 Gyrs of cooling before these conditions are reached. Consequently it could be that these types of planets are currently too hot for liquid water but will be in a far future.
\\
\\
Our results indicate that habitable conditions underneath H-He dominated atmospheres might be long-lasting and prevalent. Planets that retain their primordial atmosphere should have a relatively simple formation and evolution path compared to a secondary atmosphere. However, this is under a lot of assumptions and simplifications on the formation and evolution. \\
It is commonly assumed that habitable planets require a negative climate feedback to remain persistently habitable \cite{Walker_1981, Kasting_1993, Kopparapu_2013}. However, this work suggests that the right set of initial conditions for H$_{2}$-rich super-Earths can alone lead to persistently temperate surface conditions over Gyr timescales. There are no accurate predictions on the occurrence of super-Earth sized planets with these initial conditions, but it is likely enough that these alternatively habitable planets constitute a significant fraction of the habitable worlds in the galaxy.
\\
We made the assumption that the combination of core masses and envelope masses of these initial conditions can form. Theory \cite{Rafikov_2006} and simulations \cite{Mordasini_2018, Emsenhuber_2021b} indicate more massive envelopes are likely to form beyond the snow-line. 
Studies of observed close-in super-Earths suggest that these planets consist of H-He envelopes of several percent in mass  \cite{Lopez_2014, Rogers_2015}. Rogers \& Owen \cite{Rogers_2021} showed that the inferred post-formation H-He mass fraction peaks at $\sim$4\%, but extends well into the range needed for long $\tau_{\text{lqw}}$ (M$_{\text{env}} \, \le  \, 10^{-2.5} \, M_{\oplus}$ ). The same work showed that formation models predict too high envelope mass fractions, motivating the improvement of planet formation models. Less-massive envelopes could be the result of additional mechanisms, such as core-powered mass-loss \cite{Ginzburg_2018, Misener_2021} or collisions \cite{Liu_2015, Biersteker_2019}. We suggest that future studies should investigate whether the planetary evolution of unbound planets differs from that of bound planets, and if so, how it may affect their potential habitability. A detailed study on the formation likelihood of planets with liquid water conditions underneath a H-He envelopes that are observable at current time is beyond the scope of this research and we hope to address it in future studies. \\
Future works should investigate in detail the formation likelihood of planets with the right initial conditions. The possibility of their formation around M-dwarfs should also be considered since these stars have a prolonged pre-main-sequence with higher UV fluxes. Planets around such stars are expected to lose more of their primordial envelope and water \cite{Ramirez_2014, Luger_2015a}. Therefore, future work should consider the initial volatile abundances of such planets in order to predict the water mass fraction at different evolutionary stages  \cite{Raymond_2007, Ogihara_2009}. In addition, while a sufficient amount of water left after atmospheric escape is required, the amount of water cannot be too high in order to prevent the formation of ice layers between the silicate core and the liquid water. The exchange of nutrients would then be inhibited \cite{Alibert_2014, Noack_2016} or at least limited in the possible scenario that the ice layer is fully convective \cite{Cowan_2014}.\\
Another simplification is that the planets accrete a solar-composition envelope and remain this composition. However, processes during the evolution can increase the metallicity. This can lead to a different total greenhouse effect, with contributions of $\text{CO}_{2}$ as well as collision induced absorption by $\text{H}_{2}-\text{N}_{2}$ and $\text{H}_{2}-\text{CO}_{2}$ \cite{Wordsworth_2013, Ramirez_2014, Batalha_2015, Ramirez_2017}. We have explored the sensitivity of our result to different compositions in the Supplementary Materials.\\
Life on the type of planet described in this work would live under considerably different conditions than most life on Earth. The surface pressures in our results are on the order of 100-1000 bar, the pressure range of oceanic floors and trenches. There is no theoretical pressure limit on life and some of the extremest examples on Earth's biosphere thrive at $\sim$500 bar \cite{2016_Dalmasso}. The habitats also receive a negligible amount of direct sunlight, and therefore photosynthesis would not be an optional mechanism to provide to a metabolisms. Chemoautotrophic life on Earth \cite{Edwards_2019} would be a more likely analogue to possible life. Earth is inhabited by lifeforms that might be adapted to life underneath a H-He envelope \cite{Seager_2020}. Since chemotrophic life arose before photoautotrophic lifeforms, it can exist independently as it presumably did on Earth \cite{Wachtershauser_1990}. However it is not trivial if the emergence of life could happen on such planets in a similar way as on Earth, given that the planets we simulated spend a longer time of their evolution significantly too hot \cite{Islas_2003}. Furthermore, the advent of photosynthesis on Earth introduced a much more productive metabolism not dependent on preexisting chemical energy gradients. Consequently, planets dominated by photosynthetic life likely produce more readily observed signatures of their inhabitation.
The lack of sunlight means there would also be no temporal variation from a day-night cycle or season. 
This would lead to stable conditions (which could influence the kind of lifeforms that are best adapted\cite{Jorgensen_2007, Irwin_2020}.) Another property that can affect planetary habitability is the presence of a magnetic field \cite{Meadows_2018}. It is yet to be determined whether the planets we consider in this study are expected to generate a dynamo and we hope to address this in future research.
\\
More recent interpretations of planetary habitability do not only correspond to harboring life, but also for the ability to detect it from Earth \cite{Forget_2017}.
Currently it is still challenging to observe small planets at large radial distances, but great advancement is expected from future telescope missions. The relatively large scale-height of H-He atmospheres could make it possible for instruments like {\it JWST} and {\it Ariel} or the Extremely Large Telescope to detect and characterize biomarkers in such atmospheres \cite{Seager_2020}. While life in a $H_{2}$-dominated atmosphere environment could produce biomarkers\cite{Seager_2013b}, one challenge is that habitats that lack photosynthesis life do not produce a chemical disequilibrium, but rather destroy it by its metabolism \cite{Wogan_2020}. Future studies should predict if certain chemical disequilibria are biomarkers or anti-biomarkers for these specific habitats. Because of the importance of the system's temporal evolution, age determination by {\it PLATO} \cite{Rauer_2014} is also critical for the assessment of planetary habitability. The {\it Roman} space telescope using gravitational microlensing could detect colder exoplanets and even unbound planets \cite{Penny_2019, Johnson_2020} to constrain their population. We therefore expect that our understanding of this exoplanetary population and its potential habitability would significantly improve in the near future.\\

\begin{figure}
    \centering
    \includegraphics[width=0.7\textwidth,keepaspectratio]{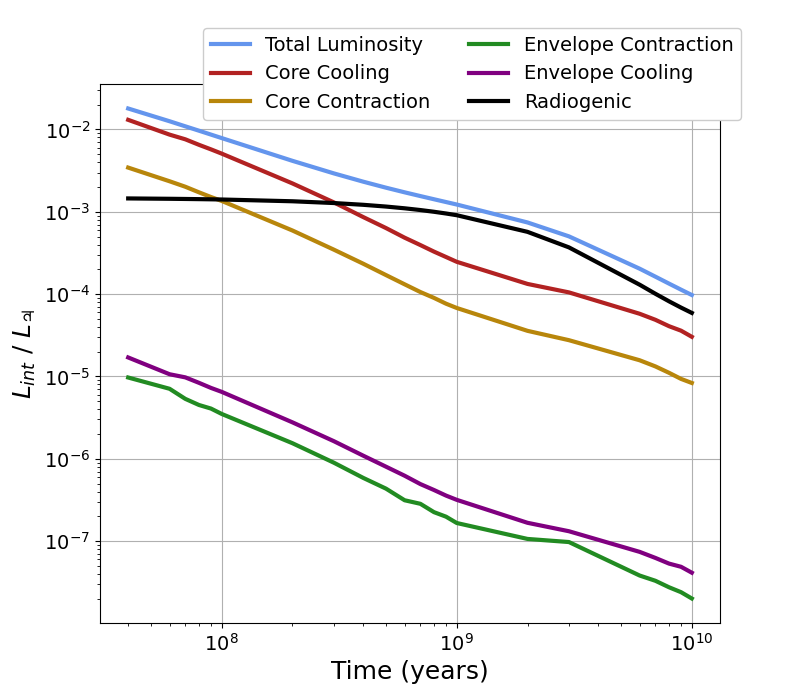}
    \caption{Evolution of the intrinsic luminosity (given in Jupiters current luminosity\cite{Guillot_2015} of 3.35$\cdot \, 10^{25}$ erg s$^{-1}$) for a planet with a core of 3 $  M_{\oplus}$ and a  0.001 $M_{\oplus}$ envelope. In the beginning of the evolution the cooling and contraction of the core are dominant components of the total luminosity. After $\sim$300 Myr the radiogenic luminosity dominates.}
    \label{fig:luminosities}
\end{figure}

\clearpage
\begin{methods}
\label{sec:method}
We simulate the gaseous envelope using a spherically symmetric model that solves the structure equations of a planet's interior assuming hydrostatic equilibrium \cite{Mordasini_2012, Jin_2014, Linder_2019}:
\\
\begin{equation} \label{eq:sse1}
\frac{\mathrm{d} m}{\mathrm{~d} r}=4 \pi r^{2} \rho 
\end{equation}
\begin{equation} \label{eq:sse2}
    \frac{\mathrm{d} P}{\mathrm{~d} r}=-\frac{G m}{r^{2}} \rho 
\end{equation}
\begin{equation}
\frac{\mathrm{d} \tau}{\mathrm{d} r}=\kappa_{\mathrm{\text{R}}} \rho 
\end{equation}
\begin{equation}
    \frac{\mathrm{d} l(r)}{\mathrm{~d} r}=0
    \label{eq:sse4}
\end{equation}
\\
This relates the local cumulative mass $m$ to the radius $r$ within the planet. $P$ and $\rho$ are the local pressure and density, respectively, and $G$ is the gravitational constant. $l$ is the local intrinsic luminosity: the energy flux going through a shell of radius $r$.  
Contributions to the luminosity due to contraction and cooling of both the envelope and the iron-silicate core (see Figure \ref{fig:luminosities}), are included as well as radiogenic heating (see Linder et al. 2019\cite{Linder_2019}). The implication of Equation \ref{eq:sse4} is that the luminosity does not vary through the gaseous envelope. This simplification is valid when the envelope has a small contribution to the total luminosity compared to the core (see Intrinsic Luminosity).
\\
The temperature gradient at high optical depth $\tau$ is represented either by the adiabatic or radiative gradient, using the Schwarzschild criterion \cite{Schwarzschild_1958, Hansen_2004}. In case of a convective region we use the adiabatic gradient:
\\
\begin{equation}
    \frac{d \, T}{d \, P} = \frac{T}{P} \left (  \frac{\text{ln} \, T}{\text{ln} \, P}  \right )_{\text{adi}}
\end{equation}
\\
At low optical depths, a double-gray atmosphere model \cite{Guillot_2010} is adapted. This atmosphere model uses the assumption that the spectral energy distribution of received stellar radiation is mostly optical light, while the outgoing radiation of the planet is mostly in the infrared. The temperature as a function of optical depth is given by:\\
\begin{equation}
\label{eq:guillot}
T(\tau)^{4}= \frac{3 T_{\mathrm{\text{int}}}^{4}} {4}\left\{\frac{2}{3}+\tau\right\}+\frac{3 T_{\mathrm{\text{eq}}}^{4}}{4}\left\{\frac{2}{3}+ \frac{2}{3 \gamma}\left[1+\left(\frac{\gamma \tau}{2}-1\right) e^{-\gamma \tau}\right]+\frac{2 \gamma}{3}\left(1-\frac{\tau^{2}}{2}\right) E_{2}(\gamma \tau) \right\}
\end{equation}
\\
The equilibrium temperature $T_{\text{eq}}$ is defined as $T_{\text{eq}} = T_{\star} \sqrt{\frac{R_{\star}}{2 a}} \left (1 - A_{\text{B}} \right )^{\frac{1}{4}}$,  where $T_{\star}$ and $R_{\star}$ are the effective temperature and radius of the star, respectively, $a$ is the distance between planet and star and $A_{\text{B}}$ is the Bond albedo fixed to Jupiter's value of 0.343. The intrinsic heat, $T_{\text{int}}$, is defined as $T_{\text{int}} = \left ( \frac{L_{\text{int}}}{\sigma_{B} 4 \, \pi r^{2}} \right )$. Here, $L_{\text{int}}$ is the total intrinsic luminosity and $\sigma_{B}$ is the Stefan-Boltzmann constant. $E_{2}(\gamma \tau)$ is a form of the exponential integral $E_{n}(z) \equiv \int_{1}^{\infty} t^{-n} e^{-zt} dt$, where $n=2$. The greenhouse effect is parametrised as the ratio between the optical opacity and the infrared opacity as $\gamma = \kappa_{\text{vis}} / \kappa_{\text{IR}}$. The level of greenhouse effect is thus contained in the parameter $\gamma$. We use a parametrisation of $\gamma$ (Table 2 in Jin et al. 2014 \cite{Jin_2014}). Since most of our planets have equilibrium temperatures that are lower than the low limit in their table ($T_{\text{eq, min}} = 260$ K), we extrapolate the table for lower $\gamma$ values. 
It should be noted that the term in Equation \ref{eq:guillot} containing the intrinsic heat becomes the dominant term in the extrapolated regime of low equilibrium temperatures, and therefore different gamma values have a small effect on the resulting temperature. 
We therefore find that the exact value of $\gamma$ in the extrapolated regime has a negligible effect on the inferred state of water underneath the atmosphere. The atmosphere then reduces to a classical Eddington grey atmosphere.\\
At high optical depths, the temperature in convectively stable regions is calculated with diffusive radiative energy transport as:
\\
\begin{equation}
    \frac{\mathrm{d} T}{\mathrm{~d} r}=-\frac{3 \kappa_{R} \rho l}{64 \pi \sigma_{\text{B}} T^{3} r^{2}}. 
    \label{eq:radiative}
\end{equation}
\\
The transition between the low optical depth regime (Equation \ref{eq:guillot}) and high optical depth regime (Equation \ref{eq:radiative}) is determined by two boundaries, between which we interpolate the temperature. These two boundaries are given by: $\tau = 10 / \sqrt{3} \gamma$ and $\tau = 100 / \sqrt{3} \gamma$. The transitions occurs at ~10 bar and ~100 bar, which means that planets which have an envelope of $\lessapprox \, 10^{-5} M_{\oplus}$ are completely in the low optical depth regime.\\
The model uses Rossland-mean opacities for a solar or a scaled-solar composition gas. One source of opacities are the molecular, grain-free tables \cite{Freedman_2014}. These are tables in temperature-density space and depend on the metallicity  [M/H], covering a range of [M/H] = 0 to 1.7 (close to 50 times solar). They include CIA that is expected to dominate at low temperatures. The lower limit on temperature for this opacity table is 75 K. Our simulations sometimes contain planets with lower temperatures in certain regions of the atmosphere (down to 50 K). In this case we extrapolate the opacities as $\propto T^{2}$ as suggested by the temperature dependency close to 75 K. To assess the consequences of the extrapolation we also repeated our simulations with a fixed opacity in the extrapolated regime. This made no difference to our final result, since the planets that we find to allow liquid water do not have a significant part of their atmosphere in the extrapolated regime.\\
We use a non-ideal equation of state \cite{Saumon_1995} for H-He \cite{Borysow_2002}. For $\text{H}_{2}\text{O}$ we use AQUA \cite{Haldemann_2020}.

\subsection{Solid Core}
Our internal structure model \cite{Mordasini_2012b} solves Equation \ref{eq:sse1} and \ref{eq:sse2} for the solid core. We assume for all cores a silicate to iron ratio of 2:1, similar to Earth, and assume there are no ices. It seems likely that planets forming beyond the snowline accrete a significant amount of water ice. On the other hand, several mechanisms dehydrate the planetary building blocks \cite{Lichtenberg_2019}. Including ices in our model has several possible consequences: First, it changes the mass-radius relationship. Since the radii of water-rich planets  are more sensitive to temperature we expect that the core radii will vary more during the evolution. Second, at a fixed total core mass, the presence of water would  reduce the silicate mantle mass and thus the amount of estimated radiogenic heating (see Importance of Intrinsic Heat). Lastly it would allow more energy to be stored in the core which is released over longer timescales, due to the fact that ice has a higher specific heat capacity than silicates. We plan to account for different planetary compositions and investigate their effect on the planetary evolution in future research.\\
The core model uses a modified polytropic EOS \cite{Seager_2007} that relates the density $\rho$ and the pressure $P$ as:\\
\begin{equation}
    \rho (P) = \rho_{0} + c P^{n}
\end{equation}\\
$\rho_{0}$, $c$ and $n$ are parameters specific to the material \cite{Seager_2007}. We use MgSiO$_{3}$ for silicates \cite{Mordasini_2012b}. The external pressure that the envelope imposes on the core is included, although this does not have a significant effect on our calculations, as the most massive envelopes considered are only 0.01 $M_{\oplus}$.

\subsection{Intrinsic luminosity}
\label{subsec:Lint}
The intrinsic luminosity of the planet and its temporal evolution are included in the simulation. We estimate the initial value of this intrinsic luminosity and its evolution in the following way.\\
First, we estimate the initial intrinsic luminosity that the planet has shortly after formation. We use an analytical fit \cite{Mordasini_2020}, which uses the results of numerous planet formation simulations. The intrinsic luminosity of planets in the formation simulation is calculated by solving the 1D structure equations, including the heating from cooling and contractions as well as accretion of gas and planetesimals. The analytical approximation yields the intrinsic luminosity as it depends on core mass, envelope mass and age of the planet. We fix the starting age of the planet at the same value as the start of our simulation, namely 20 Myr. The initial luminosity is then found by interpolating in core and envelope mass. Our simulations also show how much of the luminosity is generated in the core or the envelope (an example of which is show in Figure 4). This is also described in detail in Linder et al.\cite{Linder_2019}. It is found that luminosity contribution of the H-He envelope is much smaller than the contribution of the solid core, in agreement with earlier work \cite{Lopez_2014, Linder_2019}. This justifies the assumption of a uniform luminosity in the envelope made by Equation \ref{eq:sse4}. If a significant part of the total luminosity comes from envelope cooling and contraction, this assumption would not be valid.\\
The second contribution used to calculate the total intrinsic luminosity is based on radiogenic heating \cite{Mordasini_2012b}. We model the heat due to the decay of radioactive nucleids by assuming an initial abundance of radioactive material ($^{40}\text{K}$, $^{232}\text{Th}$, and $^{238}\text{U}$) of a chondritic composition. Then we assume that the total abundance of the radiogenic heating scales with the mass of the silicate mantle. The radiogenic luminosity becomes:
\\
\begin{equation}
    L_{\text{radio}} (t) = L_{\text{radio}, \oplus} (t) \cdot \frac{ M_{\text{mantle}}}{M_{\text{mantle}, \oplus}} 
\end{equation}
\\
Once the initial luminosity is found, its temporal evolution is given by energy conservation, i.e. by the condition that the total energy difference between two points in time is equivalent to the luminosity radiated between them \cite{Mordasini_2012}. The radioactive component evolves according to the decay time of the radioactive nucleids.\\
Figure \ref{fig:luminosities} shows an example of the evolution of the intrinsic luminosity of a 3 $  M_{\oplus}$ core and a 0.001 $M_{\oplus}$ envelope for a planet that receives a negligible amount of stellar radiation. After 300 Myr the radiogenic luminosity becomes the dominant source of intrinsic heat. Since the abundance of radioactive nucleids in other planets is unknown \cite{Wang_2020}, we also consider cases where these are a factor of ten larger/smaller, which are presented in the Supplementary Materials under Importance of Internal Heat.\\
Figure \ref{fig:PTs_in_time} shows how the atmosphere structure evolves in 3 planets from 50 Myr to 5 Gyr. All planets have a core mass of 3 $M_{\oplus}$ and an envelope of $10^{-3}$ M$_{\oplus}$. No atmosphere escape model is implemented, the evolution is due to the host star and the intrinsic luminosity. The planets at 1 and 10 AU remain too hot for liquid water conditions, while the planet at 100 AU reached liquid water conditions after 1 Gyr.

\subsection{Atmosphere loss} \label{subsec:atm_loss}
We model atmospheric loss through two distinct thermal escape models: one based on Jeans escape and one based on hydrodynamic escape.\\
Jeans escape is applicable for planets that remain in hydro-static equilibrium everywhere \cite{Opik_1961, Chamberlain_1963, Shu_1982, Tian_2015}.  The fraction of escaping particles depends on the escape velocity, particle number density and the temperature ($T_{\text{exo}}$) at the exobase. 
For the escape velocity we use the total radius and total mass of the planet and therefore assume the escape velocity at the exobase is the same as at the outer radius. We also assume a 100\% hydrogen composition for the number density and neglect that helium particles are heavier and thus have lower velocities at the same temperature when assuming hydro-static equilibrium. This results in an overestimate of the escape rate, but using an upper limit of Jeans escape serves our goal of investigating whether Jeans escape has a significant effect on liquid water duration.\\
The exosphere temperature is complex to estimate, since it is determined by the absorption of X-ray and extreme-UV, and by the composition at the exosphere \cite{Lammer_2003}. H-He dominated atmosphere are expected to have a relatively warm exosphere of $\sim$1000 K  \cite{Tian_2015}, while planets dominated by e.g. $\text{CO}_{\text{2}}$, $\text{H}_{2}\text{O}$ or $\text{N}_{2}$ should have a colder $T_{\text{exo}}$ of $\sim$300 K  \cite{Pierrehumbert_2010}. We remain agnostic about the specific exosphere temperature. Instead we put the exosphere temperature to a fixed value during the simulation ($T_{\text{exo}, \, \text{min}}$) and use a range (300 - 2000 K ). If the temperature of the outer radius of the atmosphere (where $\tau = 2/3$) is warmer than the assumed exosphere temperature, we allow the exosphere temperature to be increased to this value at any moment in the evolution. However, we find this never to be the case.\\
\\
The second escape model is based on hydrodynamic escape \cite{Jin_2014, Jin_2018}. Hydrodynamic escape is mostly driven by X-ray radiation at early times and later by extreme-UV radiation \cite{Owen_2012}. We assume X-ray dominated evaporation until a threshold of EUV flux is reached \cite{Owen_2012, Jin_2014}. The evolution of the stellar X-ray and EUV luminosity are from solar models \cite{Ribas_2005}.\\
During X-ray-dominated evaporation the loss is energy limited \cite{Watson_1981}. We estimate the mass-loss by \cite{Jackson_2012, Jin_2014}:
\\
\begin{equation}
    \dot{M}_{\text{X-ray}} = \epsilon \frac{\pi F_{\text{X-ray}} R_{2/3}^{3}}{G \, M_{\text{pl}} \, K_{\text{tide}}}
    \label{eq:Mdot}
\end{equation}
\\
$F_{\text{X-ray}}$ is the received X-ray flux, $R_{2/3}$ the radius at optical depth of 2/3, $M_{\text{pl}}$ the planet's total mass and $K_{\text{tide}}$ is an extra factor to account for a higher mass-loss when planets have their Roche-lobe boundary close to the surface \cite{Erkaev_2007}. Since we simulate relatively small planets at large distances, $K_{\text{tide}}$ has a negligible effect. In contrast to earlier work \cite{Jin_2014, Jin_2018} we now parameterise the efficiency factor $\epsilon$ by the escape velocity \cite{Wu_2019}.\\
The extreme-UV-dominated loss can also be energy limited. In this case the mass-loss is also calculated with Equation \ref{eq:Mdot}, but using the extreme UV flux and the radius where the optical depth in UV is 1\cite{Murray-Clay_2009}. When the extreme UV radiation is high, a significant amount of the heat can be lost to radiation. In this case the radiation/recombination limited (rr-limited) description \cite{Murray-Clay_2009} is used. Our model computes both the energy-limited and rr-limited mass-loss and applies the lowest value.
\subsection{Liquid water conditions}
During the simulations we determine whether there are liquid water conditions by checking the pressure and temperature at the bottom of our simulated atmosphere. We compare these to the phase-diagram of water to determine if a $\text{H}_{2}\text{O}$ water layer could be in the liquid phase \cite{Wagner_2002}. We define the duration of liquid water conditions as the time that water is permitted in the liquid phase without interruptions, referred to as $\tau_{\text{lqw}}$. In some of our results we will apply, with mentioning, an extra constraint to the liquid water conditions: that the surface temperature remains below 400 K . This is based on the observation on Earth that terrestrial live thrives best at temperatures of around 300 K and an upper limit for the chemistry of life is estimated at $\sim$400 K \cite{Takai_2008}.

\subsection{Comparison to Pierrehumbert \& Gaidos (2011) \cite{Pierrehumbert_2011} (P\&G)}
\label{subsec:comp}
To test our model in time independent calculations, we first compare our models to their results in Figure \ref{fig:compare_models}. Similarly to this work we calculate the atmosphere mass that is necessary for a surface temperature of 280 K. In addition to our own model COMPLETO we use another atmosphere model: PETITcode \cite{Molliere_2015}. PETITcode is also a 1-dimensional radiative-convective equilibrium code, but with a wavelength dependent treatment of radiative transport. It uses specific chemical abundances and assumes a chemical equilibrium. At low temperatures, however, non-equilibrium effects can play an important role in the atmosphere. It assumes a uniform gravity throughout the atmosphere.\\
The simulations of P\&G used a constant gravity of 1700 cm/$\text{s}^{2}$, without an interior heat source. To match this we set the core mass at 3 $M_{\oplus}$, which resulted in a radius of 1.32 $R_{\oplus}$. The intrinsic heat value was set to a negligible temperature (1 K  internal temperature for PETITcode, $10^{-6} \, L_{\text{jup}}$ for COMPLETO). Furthermore P\&G\cite{Pierrehumbert_2011} used a pure $H_{2}$ composition. In PETIT we set the metallicity to 0.01 times solar, so that there is predominantly hydrogen and a non-negligible fraction of helium, though we do not expect this helium to influence the opacities. In COMPLETO we use solar composition, since \cite{Freedman_2014} only provide tabulated opacities for [M / H] =0 to 1.7. The age of the host star is set at 5 Gyr.\\
The results of the PETITcode models agree very well with those of P\&G\cite{Pierrehumbert_2011}. The results are also shown with the inclusion of an intrinsic heat source of 35 K , which is the average intrinsic temperature we find for a 3 $M_{\oplus}$ planet when applying our intrinsic luminosity model. These results show that, as expected, at far semi-major axis this intrinsic heat source can reduce the amount of atmosphere that is needed to warm the surface. Quite good agreement is found inside of about 7 AU. Outside, higher pressures are found, likely a consequence of extra absorption of incoming visual light in the solar composition atmosphere relative to more H-He.

\subsection{Illustrative cases of evolving models}
Figure \ref{fig:lqc_evol_params} shows the evolution of surface pressures and temperatures for different models, and whether the pressure and temperature allow liquid water (solid lines) or not (dashed lines). We consider typical low-mass planets ($M_{\text{core}}$ 1.5 to 8 $M_{\oplus}$) with envelopes of $10^{-5}$ to $10^{-3}$ $M_{\oplus}$. The duration of continuous liquid water conditions, $\tau_{\text{lqw}}$, is therefore the integration of the (longest) solid line.\\
Figure \ref{fig:lqc_evol_params} (a) shows the effect of changing the semi-major axis. Planets relatively close to their star at 2 AU are too hot for the first 500 Myr when the combination of high irradiation (from a small semi-major axis) and a high intrinsic luminosity (from a young age) lead to high temperatures. These planets develop liquid water conditions at a later stage, while the same planet further away (at 6 AU or 10 AU) can host an ocean earlier.
The effect of changing the atmospheric mass is shown in Figure \ref{fig:lqc_evol_params} (b), where the core mass and semi-major axis are fixed. This directly influences the pressure at the bottom of the atmosphere. An envelope of $10^{-3} \, M_{\oplus}$ leads to temperatures that are too high. A smaller envelope of  $10^{-4} \, M_{\oplus}$ has liquid water conditions after 70 Myr, at high pressures and temperatures. An even smaller atmosphere of $ 10^{-5} M_{\oplus}$ has liquid water conditions at the beginning of the evolution, but then becomes too cold.
In Figure \ref{fig:lqc_evol_params}(c) we show the effect of the core mass. Changing the core mass has two different effects on our model. First of all it will result in different gravities in the atmosphere model with consequences on the radiative-convective profile. Secondly, the intrinsic luminosity depends on the core mass.\\

\subsection{Dependence on Model Parameters}
\label{subsec:modparam}
In this section we test the sensitivity of our results to the assumptions we made. We show in Figure \ref{fig:Infl_Params} how different model parameters influence duration of liquid water conditions in comparison to the nominal case presented in Figure \ref{fig:duration_noloss}.\\
The effect of including a 0.5 ice mass fraction in the core is shown in Figure 5\ref{fig:Infl_Params} (b). The nominal case assumes no ice in the core. The biggest consequence of assuming the core is half made up of ice is that it leads to an estimated mantle that is half as massive and therefore the amount of estimated radiogenics that is a factor two lower.\\
\subsection{Intrinsic Luminosity}
The second parameter we investigate is the intrinsic heat. The value and evolution of the intrinsic heat is an important, if not the main contributor to habitable surface temperatures for the planets considered in this work. We change our intrinsic heat model in two separate ways: by changing the initial luminosity due to cooling and contraction at the start of our simulation, and by changing the magnitude of the radiogenic luminosity.\\
A change of the initial luminosity (excluding the radiogenic component) has little influence on the long-term surface conditions. Planets with a more (less) initial luminosity will start with more (less) entropy. This results in more (less) efficient cooling and contractions. In our simulations, it takes around 100 Myr for planets with different initial luminosities to converge to the same luminosity. Since we find $\tau_{\text{lqw}}$ on the timescale of billions of years, it remains unaffected even when we change the initial luminosity by a factor of 4.\\
The radiogenic luminosity component is independent to the rest of the intrinsic luminosity sources. As we show in Figure \ref{fig:luminosities}, this component becomes the dominant term on long timescales and therefore we expect it to be critical on the determination of $\tau_{\text{lqw}}$. We perform our simulations, but multiply (divide) the radiogenic heat source by a factor 10. This would correspond with a planet that has 10 times more or less radioactive material as a fraction of the mantle core.\\
In Figure \ref{fig:Infl_Params} (c) and (d) we show how $\tau_{\text{lqw}}$ is influenced by changes in $L_{\text{radio}}$ of a factor 10. Up to $\sim$10 AU, $\tau_{\text{lqw}}$ seems mostly unaffected. The exception are planets with an envelope of $\sim$ $10^{-5}$ at $\sim$7 AU that are too hot for liquid water in the nominal case, but cool enough in the low $L_{\text{radio}}$ case. At distances on the order of $\sim$10-100 AU, $\tau_{\text{lqw}}$ depends on $L_{\text{int}}$. We only took into consideration the effect of a different magnitude of $L_{\text{radio}}$, which would correspond to a different total amount of radioactive particles. We did not take into account that a different composition of radioactive material could additionally lead to a different half-life.\\

\subsection{Variations in Composition}
It is likely that planets which retain their primordial atmosphere still have varying degrees of metalicity compared to a solar metallicities. Also the relative abundances of elements will vary. 
Given the uncertainty in ranges of possible exoplanet compositions, we investigate how they potentially affect our results by changing two general parameters: the greenhouse parametrization $\gamma$ in Equation \ref{eq:guillot} and the opacities.\\
The ratio between visible and thermal opacities $\gamma = \frac{\kappa_{\text{vis}}}{\kappa_{\text{IR}}}$, incorporates the greenhouse effect in the atmosphere model. Ideally, the value of $\gamma$ takes into account condensation of any molecule that absorbs in the infra-red, with contributions of all possibly relevant species such as $\text{CH}_{4}$ and $\text{CO}_{2}$.\\
In the nominal case, $\gamma$ is extrapolated from Table 2 in Jin et al. 2014 \cite{Jin_2014} from the effective temperature. Therefore the value of $\gamma$ changes in time. As discussed in Methods, the atmosphere model has a small dependency on $\gamma$ when the intrinsic temperature is significantly larger than the equilibrium temperature. This is the case for young planets and/or those that are at large semi major-axis.\\
Figure \ref{fig:Infl_Params} (e) and (f) show the effect of fixing the value of $\gamma$ to 0.01 or 0.001. It is in line with the double-grey atmosphere model that the value of $\gamma$ is an important parameter for determining $\tau_{\text{lqw}}$ at shorter distances ($>$ 3 AU), but does not effect the planets at larger distances.\\
Finally we investigate the effect of enhanced infra-red opacities in Figure \ref{fig:Infl_Params} (g). We increase the calculated opacities by a factor 10. Enhanced infra-red opacities could be the result of relatively more greenhouse gases present in the envelope, such as CH$_{4}$, NH$_{3}$ or H$_{2}$O. Our results show that enhanced infra-red opacities lead to larger surface pressures and therefore liquid water conditions would be found on planets with smaller envelopes.
In reality different greenhouse gasses would absorb at different wavelengths. To account for the contributions of specific enhanced greenhouse gasses would require a more complicated model.

\subsection{Comparison to Planet Population Synthesis}
Given the uncertainty in planet formation it is not possible to give an accurate prediction of the occurrence rate of planets that satisfy the right initial conditions leading to liquid water conditions. Nevertheless we use the New Generation Planet Population Synthesis (NG76)\cite{Emsenhuber_2021a, Emsenhuber_2021b} to compare the predicted initial conditions to those that favour long-term liquid water conditions. NG76 consists of 1000 systems, all formed around a 1 M$_{\star}$ and integrated to 100 million years. After the integration time, 34635 embryos are formed.
\\
Most planets accrete envelopes that are too big for liquid water conditions, making a potential surface temperature too hot. Nevertheless planets with the required initial conditions do occur. Out of the 34635 embryo's formed, 6851 have envelope masses of $ 10^{-6}< M_{\text{env}} / M_{\oplus}< 0.01$ and 5160 have $ 10^{-6}< M_{\text{env}} / M_{\oplus} < 0.001$. Out of these, most have a smaller core mass below 1 M$_{\oplus}$, with only 187 planets having a core mass between 1 and 10 M$_{\oplus}$. There are several reasons why the envelope mass can be overestimated in the NGPPS. First of all it applies 1-dimensional, hydrostatic models. More realistic models find reduced gas accretion because of gas exchange \cite{Cimerman_2017, Moldenhauer_2021}. The assumed grain opacities can also be an underestimation, leading to too efficient gas accretion. Other assumptions on the formation and evolution could also lead to smaller primordial envelopes than estimated in NGPPS, for example core-powered mass-loss \cite{Ginzburg_2018, Misener_2021} or collisions \cite{Liu_2015, Biersteker_2019}.\\
The fact that the NGPPS predict (too) high envelope masses is in line with the finding that planet formation theory overestimates  envelope masses in comparison to observations \cite{Rogers_2021}. It is therefore desirable to identify the missing physical/chemical processes that lead to the higher envelope masses.  We also suggest that more sophisticated planet formation and structure models are required to estimate the occurrence rate of planets with liquid water at current time.

\end{methods}

\begin{addendum}
    \item [Data Availability] Data from the simulation is public and available at\\ \href{https://github.com/mollous/Data_Liquid_Water_Conditions}{https://github.com/mollous/Data\_Liquid\_Water\_Conditions}.
    \item [Code Availability] The used code is available upon reasonable request with the corresponding author.
    \item [Acknowledgements] This work has been carried out within the framework of the National Centre of Competence in Research PlanetS supported by the Swiss National Science Foundation. The authors acknowledge the financial support of the SNSF.
    \item [Author Contributions] M. M. L. generated the data, created figures and wrote the manuscript. C.M. Provided the initial code which was modified by M. M. L. All authors analyzed the data and discussed the implications of the model assumptions, as well as the results.
    \item[Competing Interests] The authors declare that they have no competing financial interests.
    \item[Correspondence] Correspondence and requests for materials should be addressed to M.M.L. \\ ~(email: marit@ics.uzh.ch).
\end{addendum}

\section*{Figures}


\begin{figure}
    \centering
    \includegraphics[width=\hsize]{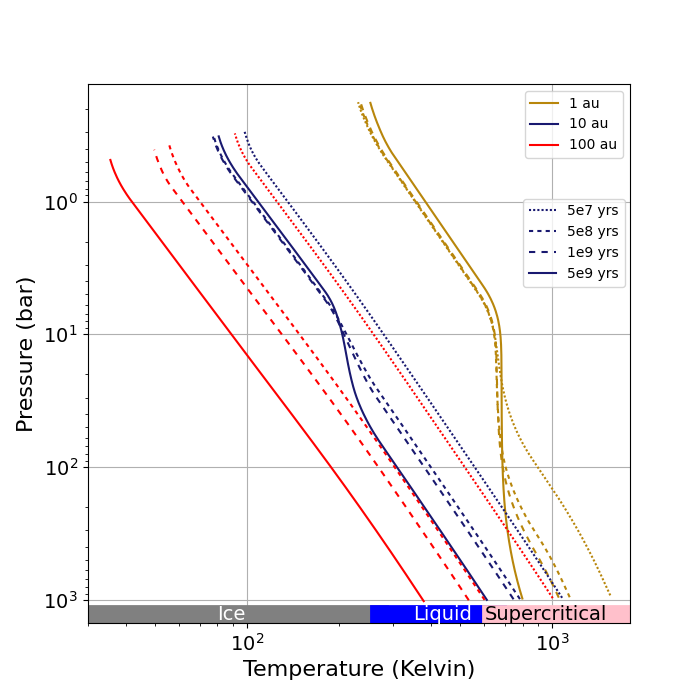}
    \caption{Evolution of the atmosphere structure of 3 planets between 50 Myr and 5 Gyr, without atmosphere loss. All planets have a core mass of 3 $M_{\oplus}$ and an envelope of $10^{-3} \, M_{\oplus}$ and are at 1, 10 and 100 AU separation from their sun-like host-star. The bar in the bottom of the figure shows the state of water for the temperature range of the figure and at a fixed pressure of about 1 kbar, the resulting surface pressure.}
    \label{fig:PTs_in_time}
\end{figure}

\begin{figure}
    \centering
    \includegraphics[width=\hsize]{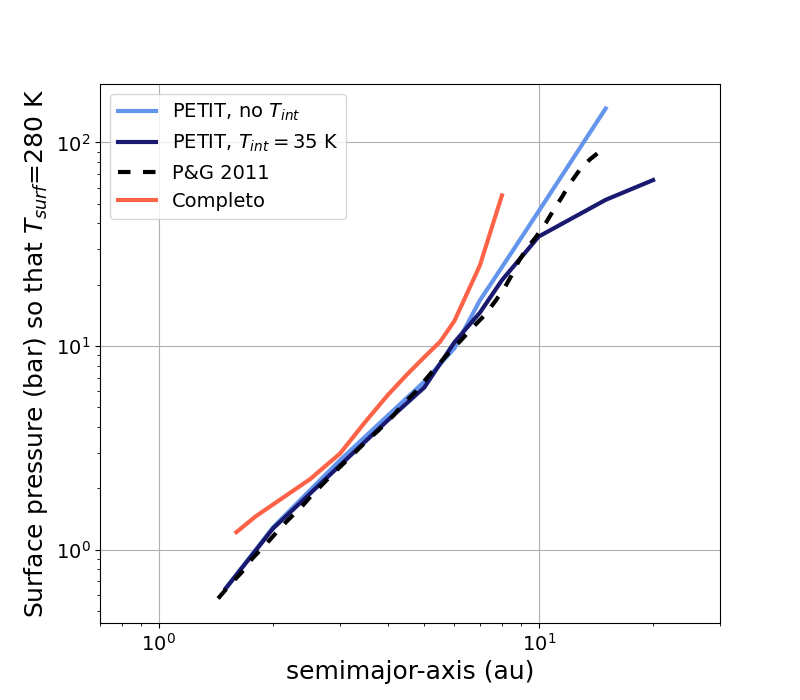}
    \caption{Comparison between the PETITcode models and the results from Pierrehumbert \& Gaidos \cite{Pierrehumbert_2011}. The figure shows the mass of a primordial, hydrogen-dominated atmosphere that is necessary for a surface temperature of 280 K, as a function of the semi-major axis around a sun-like star. The P\&G models had a surface gravity of 1700 cm/$\text{s}^{2}$, which was constant through the coordinates of the model. To match this value at the surface, the models simulate a planet with a core mass of 3 $M_{\oplus}$.}
    \label{fig:compare_models}
\end{figure}

\begin{figure}[!htb]
    \centering
    \includegraphics[width=0.5\textwidth,height=\textheight,keepaspectratio]{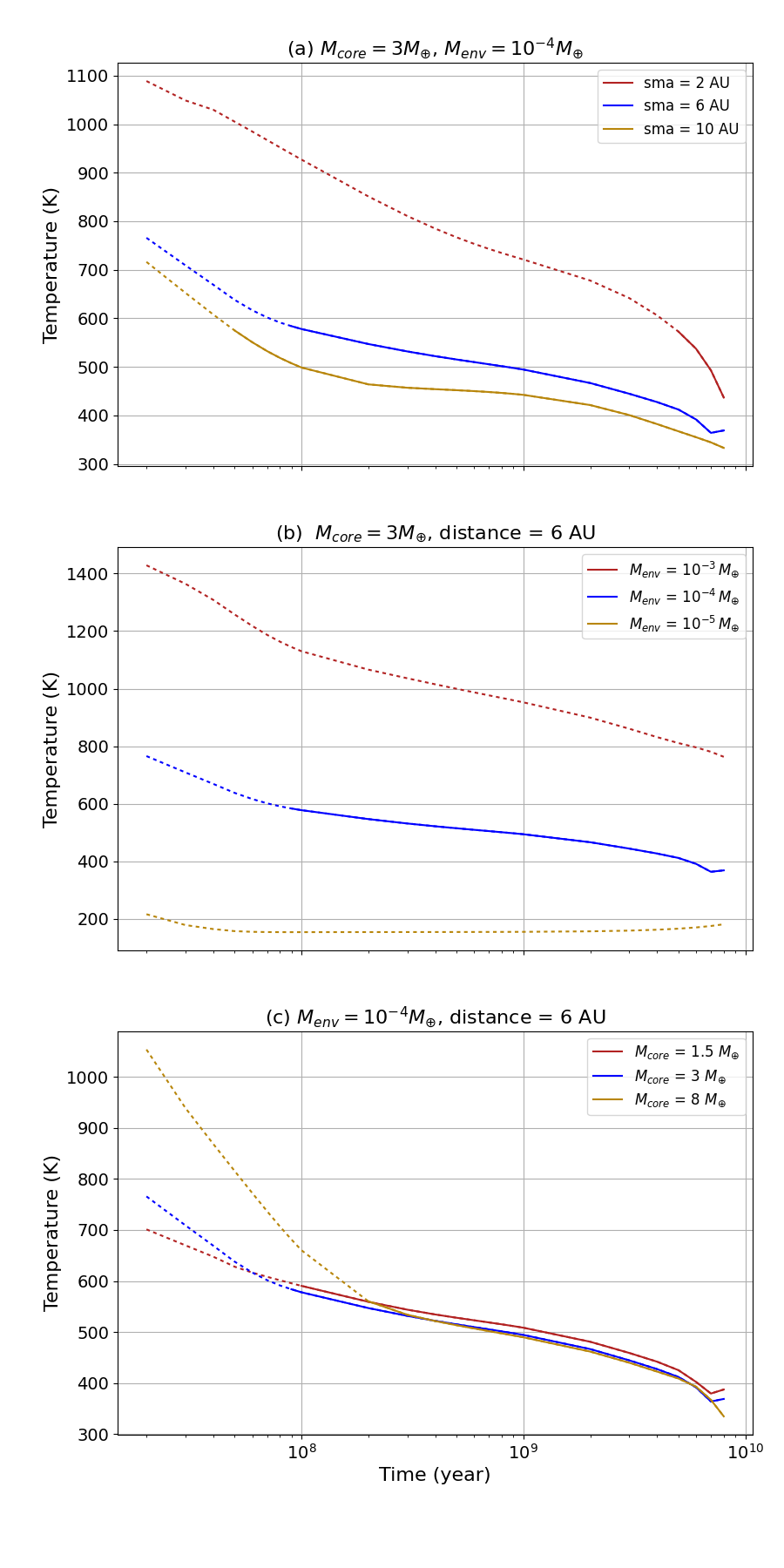}
    \caption{Evolution of the temperature at the envelope-core boundary for planets with different semimajor-axes (a), envelope masses (b) and core masses (c). The default values used are a distance of 6 AU, an envelope of $10^{-4} \, M_{\oplus}$ and a 3 $M_{\oplus}$ core. Dashed lines show when a layer of water between the envelope and core would not be in the liquid phase, solid lines show when it would. The simulations ran from 20 Myr to 10 Gyr. Our stellar model has no post Main-Sequence luminosity evolution and therefore we only consider our results until our model leaves the Main-Sequence at 8 Gyr. Atmospheric escape was not taken into account.}
    \label{fig:lqc_evol_params}
\end{figure}

\begin{figure}
    \centering
    \includegraphics[width=0.8\textwidth,height=1.3\textheight,keepaspectratio]{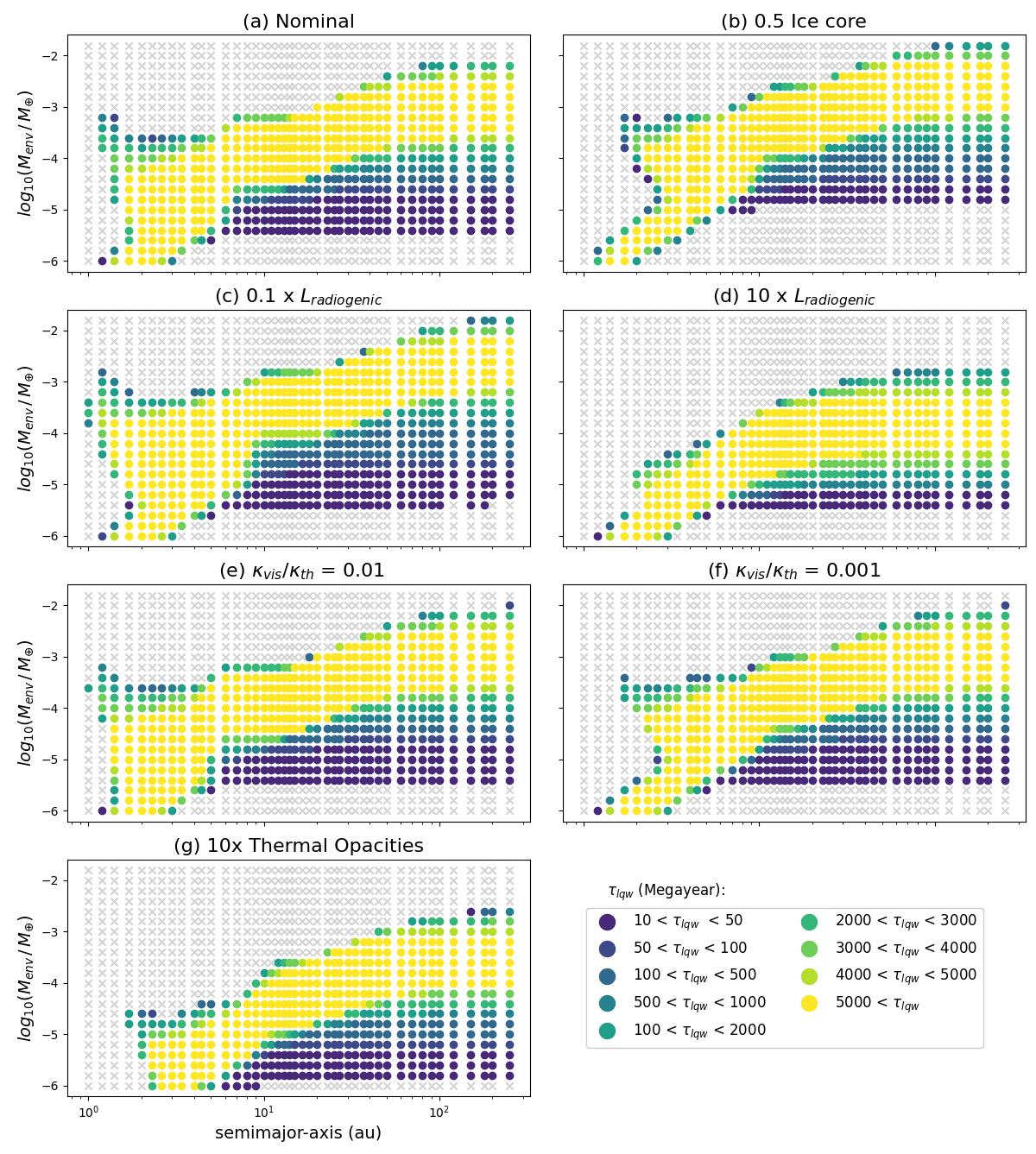}
    \caption{Comparison of $\tau_{\text{lqw}}$ for the nominal case and cases where a parameter has been changed. The semi-major axis and envelope masses are varied over a range of values on the x-axis and y-axis respectively. All cases have a total core mass of 3 M$_{\oplus}$. (a) The nominal case that is presented in Figure \ref{fig:duration_noloss}. (b) The core has a 50\% ice mass fraction, rather than 0\% in the nominal case. In (c) and (d) the amount of assumed radiogenic abundances is scaled by a factor 10 lower (higher), leading to a 10 times weaker (stronger) radiogenic luminosity. In (e) and (f) the ratio of visible to thermal opacities is fixed to values of 0.01 and 0.001. For close-in planets where the irradiation is the main source of energy these are relatively high (low, respectively) compared to the nominal case. Finally in (g) the opacities are increased by a factor 10.}
    \label{fig:Infl_Params}
\end{figure}

\clearpage
\section*{References}
\bibliographystyle{naturemag}
\bibliography{mybib}

\end{document}